\documentclass[prl,twocolumn,showpacs]{revtex4}

\usepackage{graphicx} \usepackage{dcolumn} \usepackage{bm}

\begin{document}

\title{Quantum Hall Phase Diagram  of Second Landau-level Half-filled
Bilayers: Abelian versus Non-Abelian States}

\author{Michael R. Peterson and S. Das Sarma} \affiliation{Condensed
Matter Theory Center, Department of Physics, University of Maryland,
College Park, MD 20742}

\date{\today}

\begin{abstract}
The quantum Hall phase diagram of the half-filled bilayer system in
the second Landau level is studied as a function of tunneling and
layer separation using exact diagonalization.  We make the striking 
prediction that bilayer structures would manifest two distinct branches 
of incompressible fractional quantum Hall effect (FQHE) corresponding 
to the Abelian 331 state (at moderate to low tunneling and large layer 
separation) and the non-Abelian Pfaffian state (at large tunneling and 
small layer separation).  The observation of these two FQHE branches and 
the quantum phase transition between them will be compelling 
evidence supporting 
the existence of the non-Abelian Pfaffian state in the second Landau level.
\end{abstract}


\maketitle

The fractional quantum Hall effect (FQHE) in the second Landau level
(SLL) at filling factor $\nu=5/2$~\cite{willett} presents an
amazing confluence of ideas from condensed matter physics, conformal
field theory, topology, and quantum computation.  In particular, the
fundamental nature of the experimentally observed 5/2 FQHE, whether an
exotic spin-polarized non-Abelian incompressible paired state or a
more common Abelian incompressible paired state, has remained an
enigma for more than 20 years.  Although theoretical and numerical
work indicates that the 5/2 FQHE belongs to a
non-Abelian Pfaffian universality class~\cite{tqc-rmp}, there is scant experimental
evidence supporting this conclusion~\cite{dean}.  Since the question
of non-Abelian-ness (or not) of the experimental 5/2 state is of
profound importance beyond quantum Hall physics~\cite{tqc-rmp}, 
it is useful to contemplate novel
situations where the nature of the 5/2 state will manifest itself in a
dramatic, but hitherto unexplored, manner.  In this Letter, we make a
specific suggestion for such a possibility by showing theoretically
that a bilayer system may provide important new insight into the
nature of the 5/2 FQHE by allowing for a novel quantum phase
transition between an Abelian and non-Abelian 5/2 FQHE as system
parameters are varied.  The system we have in mind could either be a
true bilayer double-quantum-well structure or a single
wide-quantum-well which manifests effective bilayer behavior.  We show
that tuning the inter-layer tunneling and/or the layer separation
would lead to a transition between the
Abelian and the non-Abelian SLL 5/2 FQHE, which should be
\emph{easily} observable experimentally in standard FQHE 
transport experiments.

We consider the FQHE in spin-polarized bilayer quantum wells with
non-zero tunneling at total filling factor $\nu=5/4+5/4=5/2$, i.e.,
the half filled SLL with each layer having one quarter SLL filling.
The best known variational $N$ electron wavefunctions describing the physics
of this system are the paired Abelian Halperin 331 (331)
state~\cite{halperin-331}
\begin{eqnarray}
\Psi_{331}=\prod_{i<j}^{N/2}(z^R_i-z^R_j)^3\prod_{i<j}^{N/2}(z^L_i-z^L_j)^3\prod_{i,j}^{N/2}(z^R_i-z^L_j)
\end{eqnarray}
which is known~\cite{r1} to be a good description of two-component FQHE
states at half-filling and the non-Abelian Moore-Read Pfaffian state (Pf)~\cite{mr-pf}
\begin{eqnarray}
\Psi_{\mathrm{Pf}}=\mathrm{Pf}\left(\frac{1}{z_i-z_j}\right)\prod_{i<j}^{N}(z_i-z_j)^2
\end{eqnarray}
which is a good description of one-component FQHE
states~\cite{tqc-rmp} at half filling ($z=x-iy$ is the electron position and the
superscript $R(L)$ indicates the right(left) layer).  

The interesting possibility, which we validate through numerical calculations, is that
by tuning the inter-layer tunneling, one can go from an incompressible
FQH state at 5/2 described by a two-component 331 bilayer state in the
weak-tunneling regime to a one-component Pf state in the
strong-tunneling regime.  Although a two-component 331 bilayer FQH
state has been observed~\cite{eisenstein-prl,shayegan-prl} at
$\nu=1/2$ in the lowest LL (LLL), the corresponding one-component
strong-tunneling state turns out to be the non-FQHE compressible composite
fermion sea in the LLL.  In the SLL, we establish the possibility of
observing the novel Abelian-to-non-Abelian quantum phase transition
(with no compressible state in between) in the bilayer $\nu=5/2$
situation.  Our predictions can be experimentally tested in bilayer
structures or in thick single-layer structures (where the
self-consistent field from the electrons produces effective two-component
behavior).

Theoretically, the system is parameterized using three variables: the
width $w$ of a single layer (both layers having the same width), the
separation $d$ between the two layers (by definition $d>w$ since $d$ is
the distance between the center of the two layers), and the inter-layer tunneling
energy $\Delta$ (i.e., the symmetric-antisymmetric energy gap).  The most
natural theoretical approach is to use the symmetric-antisymmetric
basis where $c_{mS}=(c_{mR}+c_{mL})/\sqrt{2}$ and
$c_{mA}=(c_{mR}-c_{mL})/\sqrt{2}$ destroy an electron in the symmetric
($S$) and antisymmetric ($A$) superposition states, respectively, where
$m$ is angular momentum and $c_{mL(R)}$ destroys an electron in the
left(right) quantum well.  $S(A)$ can be considered to be an effective
pseudospin index for the bilayer system.  In this basis, the
Hamiltonian of spin-polarized interacting electrons in a bilayer
quantum well system entirely confined to a single Landau level (all completely 
filled LLs are considered inert) with tunneling is given by
\begin{widetext}
\begin{eqnarray}
H&=&\frac{1}{2}\sum_{\{m_i,\sigma_i=A,S\}}\langle
m_1\sigma_1,m_2\sigma_2|V|m_4\sigma_4,m_3\sigma_3\rangle
c^\dagger_{m_1\sigma_1}c^\dagger_{m_2\sigma_2}c_{m_4\sigma_4}c_{m_3\sigma_3}
-\frac{\Delta}{2}\sum_{m}(c^\dagger_{mS}c_{mS} -
c^\dagger_{mA}c_{mA})\;.
\end{eqnarray}
\end{widetext}
The intra-layer Coulomb potential energy between two electrons is
$V=(e^2/(\epsilon l)/\sqrt{r^2 + w^2}$ (we use the Zhang-Das
Sarma~\cite{zds-potential} potential to model the single layer
quasi-2D interaction) and the inter-layer potential energy is
$V=(e^2/(\epsilon l))/\sqrt{r^2 + d^2}$ (both $d$ and $w$ are given in
units of the magnetic length $l=\sqrt{\hbar c/eB}$ where $e$ is the 
electron charge and $B$ the magnetic field strength).  In other words,
we solve the bilayer interacting Hamiltonian in the $d$-$w$-$\Delta$
parameter space in the $S_z$ ($z$-component of the pseudospin operator) basis.  
All lengths and energies are given in units of the magnetic length from now on.  
We use the magnetic length $l$ and the corresponding Coulomb 
energy $e^2/\epsilon l$ ($\epsilon$ is the dielectric constant of the host 
semiconductor) as the unit of length and energy, 
respectively, throughout (including figures).

We exactly diagonalize $H$ in the SLL, for finite $N$, utilizing 
the spherical geometry where the electrons are confined to
a spherical surface of radius $R=\sqrt{N_\phi/2}$, $N_\phi/2$ is the
total magnetic flux piercing the surface ($N_\phi$ is an integer
according to Dirac), and the filling factor in the partially 
occupied SLL is $\lim_{N\rightarrow\infty}N/N_\phi$.  Incompressible ground states
(those exhibiting the FQHE) are uniform states with total angular momentum
$L=0$ and a non-zero excitation gap.  We are concerned with the Pf and
331 variational states for total $\nu=5/4+5/4=5/2$ so $N_\phi=2N-3$.  Throughout, 
we consider $(N,N_\phi)=(8,13)$ and mention that due to the pseudospin
component present in the bilayer problem the Hilbert space is much
larger (more than $10^5$ states for $N=8$ and over
$7\times10^6$ for the next largest system at $N=10$) than typical
one-component FQHE systems.  The $N=6$ electron system is aliased with 
$2+3/7$ filled LLs leading to possible ambiguities.  $N=8$ should be
adequate for a qualitative and semi-quantitative understanding of the
physics.

We investigate this system with the usual probes used in theoretical
FQHE studies by calculating: (i) wavefunction overlap between
variational ansatz (Pf and 331) and the exact ground state--an overlap
of unity(zero) indicating the physics is(is not) described by the
ansatz; (ii) expectation value of the $z$-component of the pseudospin
operator $\langle S_z\rangle$ of the exact ground state--a value of
zero($N/2$) indicating the ground state to be two-(one-)component; and
(iii) energy gap (provided the ground state 
is a uniform state with $L=0$)--a non-zero excitation gap indicating a possible FQHE
state.  In other words, we ask: 
(i) What is the physics (i.e., 331 or
Pf)?; (ii) Is the system one- or two-component?; (iii) Will the
system display FQHE (i.e., is the system compressible or
incompressible)?

\begin{figure}[t]
\begin{center}
\mbox{\includegraphics[width=6.75cm,angle=0]{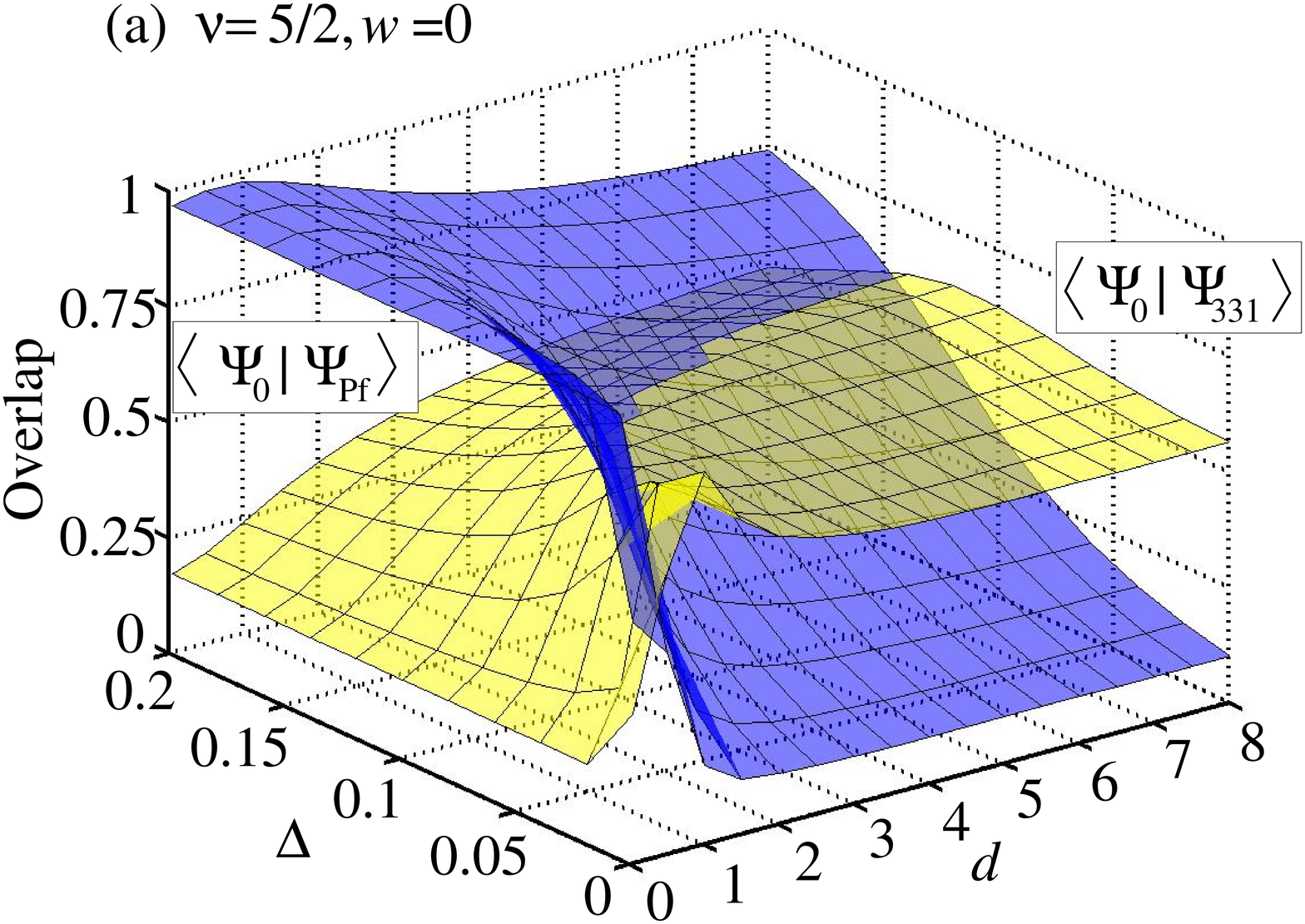}}\\
\mbox{\includegraphics[width=6.75cm,angle=0]{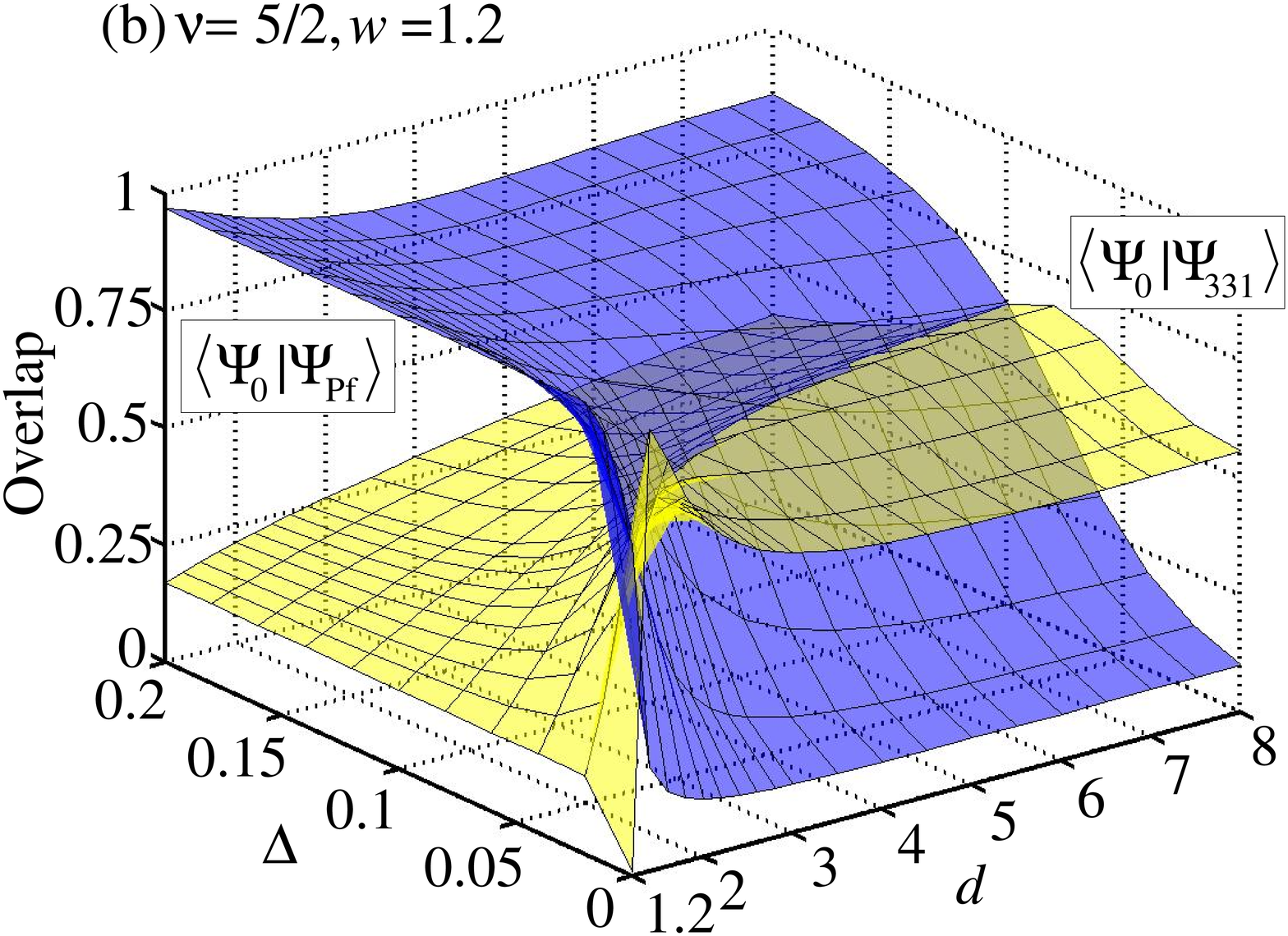}}
\end{center}
\caption{(Color online) Wavefunction overlap between the exact ground
state $\Psi_0$ and the Pf state $\Psi_{\mathrm{Pf}}$ (blue (dark
gray)) and 331 state $\Psi_{331}$ (yellow (light gray)) as a function
of distance $d$ and tunneling amplitude $\Delta$ for the half-filled 
second LL with $N=8$ electrons and width (a) $w=0$ and (b) $w=1.2$ ($d>w$ necessarily).}
\label{fig1}
\end{figure}

\emph{(i) What is the physics?}  The calculated
wavefunction overlap between the exact ground state $\Psi_0$ of $H$
and the two appropriate candidate variational wavefunctions ($\Psi_{\mathrm{Pf}}$ 
and $\Psi_{331}$) as a function of
distance $d$ and tunneling energy $\Delta$ is shown in Fig.~\ref{fig1}.  
First we focus on the
situation with zero width $w=0$ (Fig.~\ref{fig1}a) and concentrate on
the overlap with $\Psi_\mathrm{Pf}$.  In the limit of zero tunneling
and zero $d$ the overlap with $\Psi_\mathrm{Pf}$ is small--approximately 0.5
(unclear from the figure).  However, only weak tunneling is required
to produce a state with a sizeable overlap of approximately $\sim0.96$
and as the tunneling is increased this overlap remains large and
approximately constant.  For $d\neq 0$, in the weak tunneling limit,
the overlap with $\Psi_\mathrm{Pf}$ decreases drastically.  Adding moderate to strong
tunneling we obtain a sizeable overlap, decreasing gently in the
large $d$ limit.  This shows that the strong-tunneling (one-component)
regime is well described by the Pf state.

Next we consider the overlap between $\Psi_0$ and $\Psi_{331}$.  In
the zero tunneling limit, as a function of $d$, we find the overlap
starts small, increases to a moderate maximum of
$\approx 0.80$ at $d\sim1$ before achieving an essentially constant
value of $\approx0.56$.  For $d>4$ the overlap remains relatively
constant and slowly decreases as the tunneling is increased (in fact,
there is a slight increase in the overlap to $\sim0.6$ for a region of
positive $d>4$ and $0.1\lesssim\Delta\lesssim
0.15$).  Thus, the weak-tunneling (two-component) regime is well 
described by $\Psi_{331}$.

In Fig.~\ref{fig1}b we consider the finite layer width situation (i.e.
$w\neq 0$) which may enhance the overlap~\cite{mrp-prl-prb} with
paired FQH states in the SLL.  Indeed, if we consider zero tunneling
and $w>0$ (and necessarily $d>w$) we find that the overlap between
$\Psi_0$ and $\Psi_{331}$ can be increased significantly to $\sim0.94$
by using $w=1.2$ and $d=1.4$.  There is a region of large overlap
for $1.1<d/w<2.5$ (provided $w\lesssim2$).  The overlap
with $\Psi_{331}$ increases for finite $w$ compared with the $w=0$ case, while
in the strong tunneling limit, the overlap decreases with increasing
$w$.  The overlap with $\Psi_\mathrm{Pf}$ is also affected; namely, in
the strong tunneling regime the overlap increases for $d\neq 0$ and
the amount of tunneling required to increase the overlap decreases.

\begin{figure}[t]
\begin{center}
\mbox{\includegraphics[width=6.25cm,angle=0]{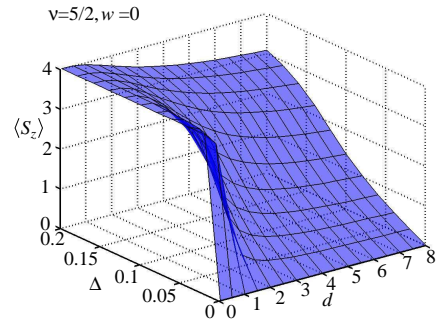}}
\end{center}
\caption{(Color online) $\langle S_z\rangle$ as a function of distance
$d$ and tunneling amplitude $\Delta$ 
for $\nu=5/2$, $N=8$ electrons, and $w=0$.  (The $w\neq 0$ case is 
nearly identical.)}
\label{fig2}
\end{figure}

\emph{(ii) Is the system one- or two-component?}  Our conclusion so
far, based on overlap calculations, is completely consistent with the calculated
expectation value of the $z$-component of the pseudospin operator,
$\langle S_z\rangle$ which is essentially the order parameter
describing the one-component to two-component transition, i.e.,
$\langle S_z\rangle\approx N/2$ describes a one-component phase
whereas $\langle S_z\rangle\approx 0$ describes a two-component phase.
Fig.~\ref{fig2} shows the calculated $\langle S_z\rangle$ as a
function of $d$ and $\Delta$ with $w=0$ (the $w=1.2$ result is
almost identical).  For $d=0$ and $\Delta>0.1$ the
system is essentially one-component as evidenced by $\langle
S_z\rangle\sim N/2$ and it is expected that the Pf would provide a
good physical description as it indeed does (cf. Fig.~\ref{fig1}).  In
the region where $d\neq 0$ and $\Delta$ is small, the system is
essentially two-component as evidenced by $\langle S_z\rangle\sim 0$
and one would expect that the 331 state would provide a good physical
description as it indeed does.

\begin{figure}[t]
\begin{center}
\mbox{\includegraphics[width=7.25cm,angle=0]{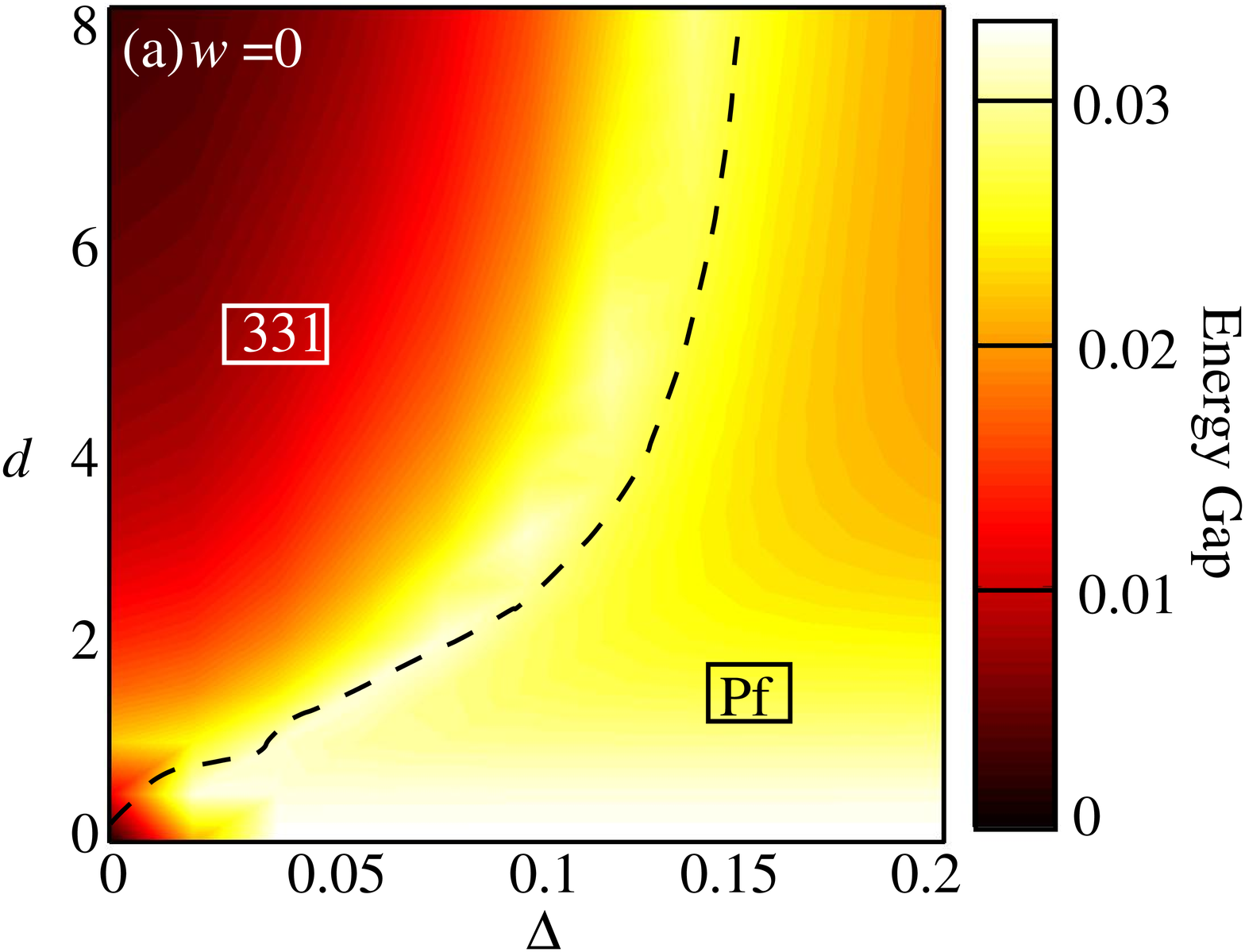}}\\
\mbox{\includegraphics[width=7.25cm,angle=0]{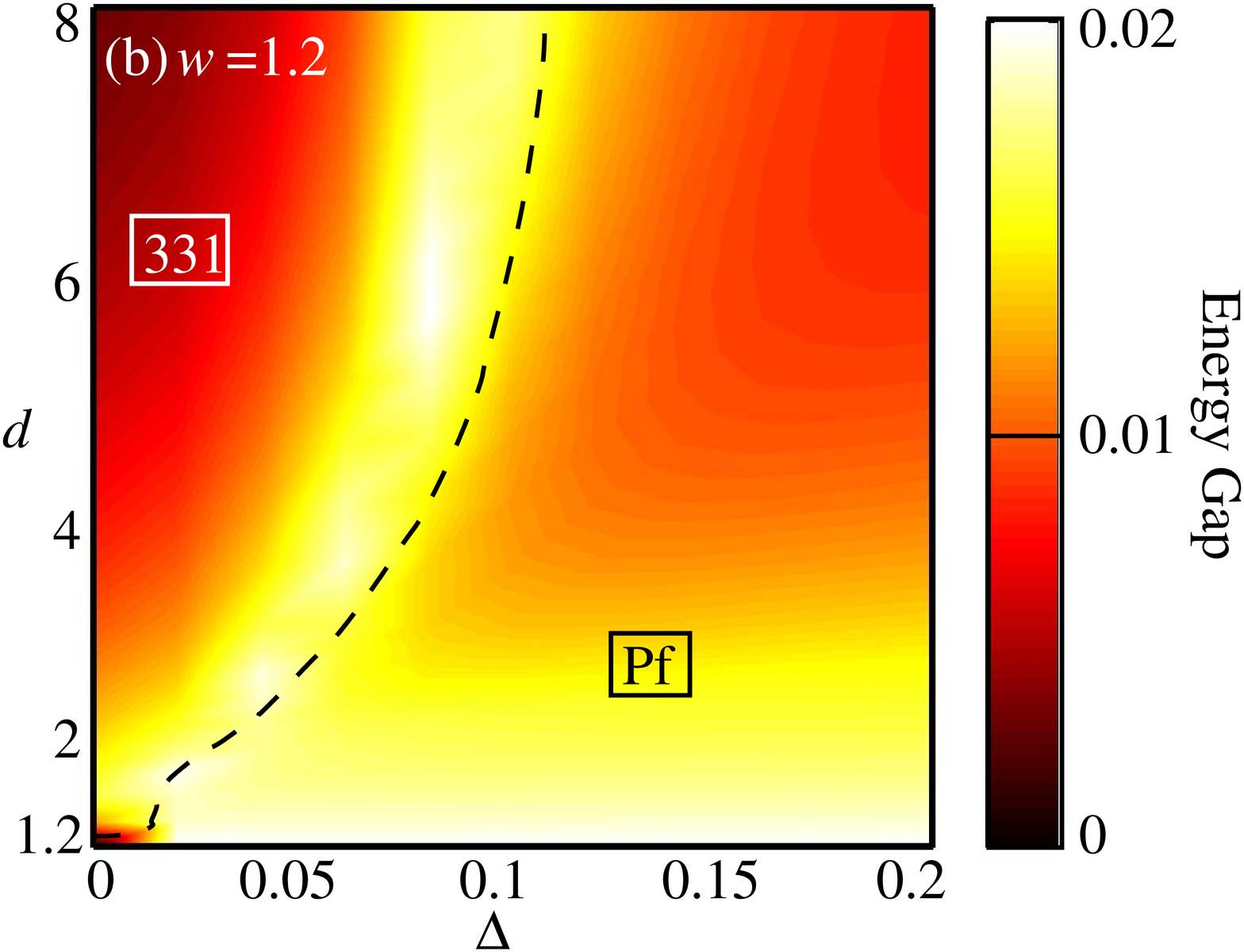}}
\end{center}
\caption{(Color online) Bilayer quantum phase diagram showing contour plots 
of the FQH SLL energy gap for $\nu=5/2$ as a function 
of layer separation $d$ and tunneling energy $\Delta$ for
the second LL, $N=8$ electrons, and (a) $w=0$ and 
(b) $w=1.2$.  The black dashed line separates the upper(lower) 
region where the overlap with 331(Pf) is larger.}
\label{fig3}
\end{figure}

\emph{(iii) Will the system display the FQHE?}  Wavefunction overlap
and pseudo-spin are only two properties that elucidate the physics.
Another property is the energy gap above the $L=0$ ground state in
the excitation spectra; a crucial characteristic that determines
whether the system would display the FQHE at all.  In
Fig.~\ref{fig3}a(b) we show the energy gap, defined as the difference
between the first excited and ground state energies at constant
$N_\phi$, as a function of $d$ and $\Delta$ for the SLL system for
$w=0$($w=1.2$).  In the $d=0$ and $\Delta=0$ limit the system is gapless, 
however, a gap suddenly appears when the SU(2) symmetry is
broken explicitly for $d\neq 0$.  Most interestingly, the gap (other
than the SU(2) symmetric point) is always positive throughout the
parameter space.  Thus, if one starts in a region described by $\Psi_{331}$,
then one can smoothly move into a region described by $\Psi_{\mathrm{Pf}}$ without
the gap closing.  The phase transition between two topologically
distinct FQHE states, the Abelian 331 and non-Abelian Pf state, is
continuous at $w=0$ with no compressible state in between.  This is an
unexpected interesting feature of our results.  

The other notable feature is the large energy gap in the parameter
region where the system crosses over from a Pf to a 331 state.
However, we stress that the region with the \emph{largest} gap is the one with $d=0$
and finite $\Delta$ described by $\Psi_\mathrm{Pf}$.  It makes
intuitive sense that the largest gap would be when $d=0$ since any
form of finite thickness trivially lowers the energy by softening the
Coulomb interaction, but this is in stark contrast to the results (not
shown) in the LLL where the $d\approx 0$ region has a lower gap than
the ``crossover'' region which has the largest gap.  The reason for
this quantitative difference between the LLL and SLL regarding the
location of the optimal gap in the parameter space, i.e. at the
crossover regime for the LLL and at $d=0$ for the SLL, is a matter of
quantitative details and arises from the different behaviors of the
Coulomb pseudopotentials in the two different Landau levels.  We note from
investigating the energy gap that at $\nu=5/2$ the Pf one-component
FQHE state would be more robust, i.e, would survive to higher
temperatures and larger disorder strength, than the 331 two-component FQHE
state where the gap, in that parameter region, is lower in part
because the Coulomb interaction has been softened by the finite
distance required to stabilize it.  This may be one simple way to
experimentally distinguish the two states in the bilayer system.

Finally, we present the quantum phase diagram for the bilayer system 
at $\nu=5/2$ in Fig.~\ref{fig3}.  This 
phase diagram is created by identifying the 331 and 
Pf phases as the regions in the parameter space where the 
overlap with $\Psi_0$ is larger, cf. Fig.1.  In general, the 
one-component 331 phase (the upper left region) has
a weaker SLL FQHE ($\nu=5/2$) than that of the two-component Pf (the lower
right region).  This should be contrasted with the corresponding LLL case
where the 331 phase is the \emph{only} observed bilayer $\nu=1/2$ FQHE
state~\cite{eisenstein-prl,shayegan-prl,luhman-2009} and the Pf state loses
out to the 331 state~\cite{he-papic-nomura}.  Figure~\ref{fig3}, which is the
most important prediction of our work, shows that in the SLL bilayer
system, both the 331 and Pf states should be visible, and in a realistic
finite thickness system (Fig.~\ref{fig3}b), the FQHE gap would become
very small as one goes from the 331 to the Pf regime.  In fact, it is conceivable 
that in the finite $w$ system (Fig.~\ref{fig3}b) the quantum phase transition 
between the 331 and the Pf phase can only occur by going through an 
intermediate compressible phase.  The
experimental observation of two strong FQHE regimes, like 
those shown in Fig.~\ref{fig3}b, with the FQHE gap being largest along the two ridges in
the phase diagram as $d$ and $\Delta$ are varied should be a
spectacular verification of our theory and strong evidence for the existence 
of a Pf FQHE in the SLL.  

To summarize, we predict the existence of both
the two-component 331 Abelian (at intermediate to large $d$ and small
$\Delta$) and the one-component Pfaffian non-Abelian (at small $d$ and
intermediate to large $\Delta$) $\nu=5/2$ SLL FQHE phase in bilayer
structures.  The observation of these two topologically distinct
phases, one (Pf) stabilized by large inter-layer tunneling and the
other (331) stabilized by large inter-layer separation would be a spectacular
verification of the theoretical expectation that bilayer structures
allow quantum phase transitions between topologically trivial and
non-trivial paired even-denominator incompressible FQHE states.  
The direct experimental observation of our predicted 
``two distinct branches'' of SLL 5/2 FQHE in bilayer structures, as shown 
in Fig.~\ref{fig3}b, will be compelling evidence for the existence of the $\nu=5/2$ 
non-Abelian Pf state.

We acknowledge support from Microsoft Project Q.  MRP thanks Kentaro
Nomura for a helpful discussion.

\end{document}